\documentclass[aps,prl,twocolumn,showpacs,amsmath,amssymb,amsfonts, long,preprintnumbers,superscriptaddress]{revtex4}
\usepackage{graphicx}

\newcommand{\Msun}{M_{\odot}}

\begin{document}

\title{Constraining the Axion-Photon Coupling with Massive Stars}
\author{Alexander Friedland}
\email{friedland@lanl.gov}
\affiliation{Theoretical Division, T-2, MS B285, Los Alamos National Laboratory, Los Alamos, NM 87545, USA}

\author{Maurizio Giannotti}
\email{mgiannotti@mail.barry.edu}
\author{Michael Wise}
\email{mwise@mail.barry.edu}
\affiliation{Physical Sciences, Barry University,
11300 NE 2nd Ave., Miami Shores, FL 33161, USA}

\date{October 3, 2012}

\preprint{LA-UR-12-25074}


\begin{abstract}
We point out that stars in the mass window $\sim8-12\Msun$ can serve as sensitive probes of the axion-photon interaction, $g_{A\gamma\gamma}$. Specifically, for these stars axion energy losses from the helium-burning core would shorten and eventually eliminate the blue loop phase of the evolution. This would contradict observational data, since the blue loops are required, {\it e.g.}, to account for the existence of Cepheid stars. Using the  MESA stellar evolution code, modified to include the extra cooling, we conservatively find
$g_{A\gamma\gamma} \lesssim 0.8 \times 10^{-10}$ GeV$^{-1}$, which compares favorably with the existing bounds. 
\end{abstract}


\maketitle

\section{Introduction}
For a particle physicist, stellar interiors represent extremely hermetic  detectors, sensitive to very rare processes. 
For example, the $\gamma^{\ast}\rightarrow\nu\bar\nu$ decay \cite{Adams:1963zzb} measurably drains energy from the core of a 
red giant star, even though the probability of this decay to occur between successive interactions of the plasmon $\gamma^{\ast}$ is only $\sim 10^{-26}$.
Importantly, the rate of the energy drain cannot be too different from the Standard Model (SM) prediction, allowing one to constrain neutrino properties
~\cite{Bernstein:1963qh,Sutherland:1975dr,Dicus:1976ra} (see \cite{Raffelt:1990yz} for further references).
%
The same argument extends to new physics scenarios with light, weakly interacting particles \cite{Sato:1975vy}.
Numerous examples include majorons \cite{Georgi:1981pg}, light supersymmetric partners \cite{Fukugita:1982eq}, novel baryonic or leptonic forces \cite{Grifols:1986fc}, and more recently unparticles \cite{Hannestad:2007ys} and extra-dimensional photons \cite{Friedland:2007yj}.
A particularly compelling scenario is furnished by the axion~\cite{Dicus:1978fp,Sato:1978vy}, which is the subject of this letter. Below, we reexamine the astrophysical implications of the axion-photon coupling and point out that, contrary to the standard lore, stars with masses $\sim8-12\Msun$ are very sensitive to it.

The axion arose from a proposal to account for the absence of CP violation in the strong interactions (QCD) \cite{Peccei:1977hh,Peccei:1977ur,Weinberg:1977ma,Wilczek:1977pj}. The SM QCD Lagrangian admits a CP-violating $G\tilde{G}$ term, which, if present, would impact physical amplitudes through nonperturbative effects \cite{Hooft:1976kx,Hooft:1976fk,Callan:1976je,Jackiw:1976pf}. In particular, one may expect the neutron to have a large electric dipole moment ~\cite{Baluni:1978rf,Crewther:1979pi}, contrary to observations~\cite{Baker:2006ts}. The axion proposal addresses this by promoting the coefficient of the $G\tilde{G}$ term to a dynamic field, which is constructed to be the Goldstone component of a $U(1)$ field. The corresponding broken symmetry (Peccei-Quinn) is anomalous, hence the Goldstone couples to the SM fields, particularly the pion, and gains a small potential. This potential dynamically drives the axion field to the CP conserving vacuum, solving the problem.

Being a pseudo-Goldstone boson, the axion can be light enough to be produced in stars. More precisely, the axion mass $m_{A}$ and decay constant $f_{A}$ are related to those of the pion, as $m_{A}f_{A}\simeq m_{\pi}f_{\pi}$, or \cite{Weinberg:1977ma,Bardeen:1977bd,Kolb:1990vq}
\begin{equation}
(m_{A}/1\mbox{ eV})(f_{A}/10^{7}\mbox{ GeV})\simeq0.6. 
\label{eq:massfA}
\end{equation}
Below, we will be interested in axion emission from He-burning stellar cores, which have temperatures $\sim 10^{4}$ eV. Eq.~(\ref{eq:massfA}) then tells us that for $f_{A}$ above the weak scale the axion is indeed light enough to be thermally produced.

The high scale of $f_{A}$ also ensures the second condition:  axions interact weakly enough to free-stream out of stellar cores. The couplings of the axion field $\phi_{A}$ to axial SM currents $J_{f}^{\mu}=\bar\Psi_{f}\gamma^{\mu}\gamma_{5}\Psi_{f}$, and to photons are both suppressed by $f_{A}$: ${\cal L} \in C_{f}f_{A}^{-1} J_{f}^{\mu} \partial_{\mu} \phi_{A} + C_{\gamma}\alpha/(8\pi f_{A})\phi_{A}F_{\mu\nu}\tilde F^{\mu\nu}$. It is easy to verify that axions emitted from $He$-burning stellar cores do not reinteract.

\begin{figure}[bt]
  \includegraphics[angle=0,width=\columnwidth]{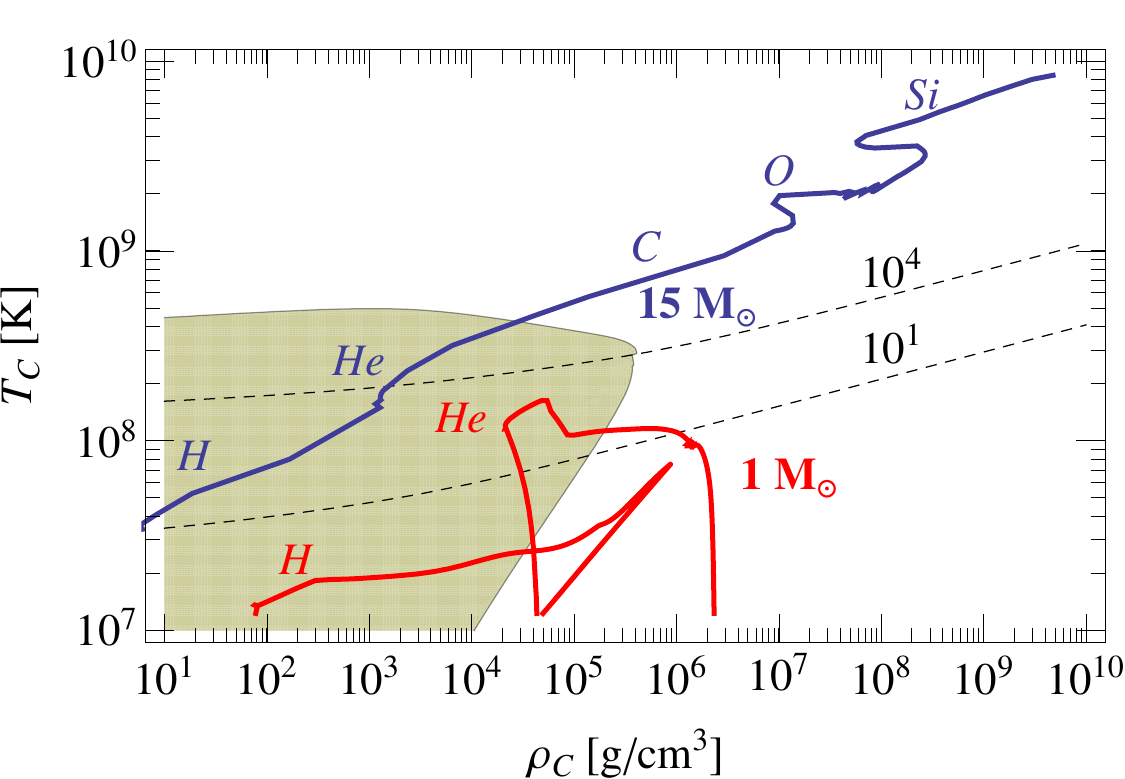}
  \caption{The Figure shows why $He$ burning is optimal for probing the $A\gamma\gamma$ coupling. The curves trace the evolution of the central density $\rho_{C}$ and temperature $T_{C}$ in stars of 1 $M_{\odot}$ {\it (bottom, red)} and 15 $M_{\odot}$ {\it (top, blue)}. Different burning stages are labeled. The shaded region shows the range of conditions, for which axion emission with $g_{10}=1$ contributes at least 90\% of the non-photon energy loss. The dashed isocontours correspond to loss rates of $10^{1}$ and $10^{4}$ erg/g/s, as labeled.}
  \label{fig:RhoThistories}
\end{figure}   

\begin{figure*}[t]
  \includegraphics[angle=0,width=\columnwidth]{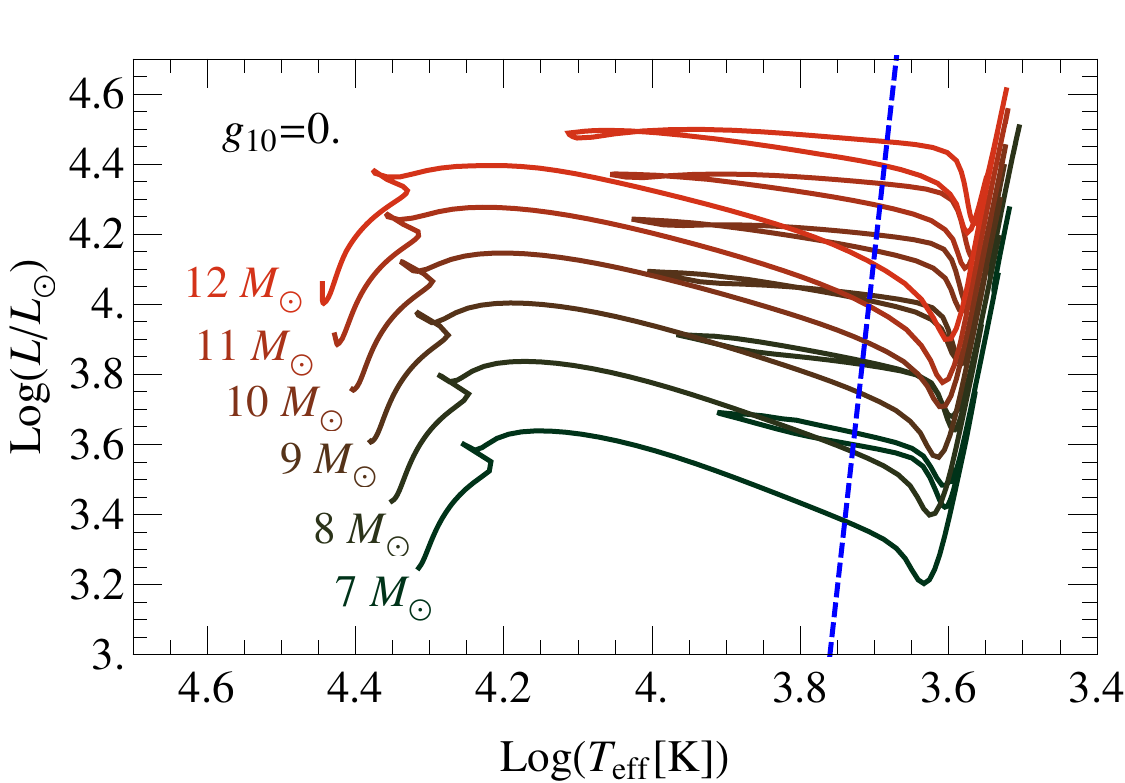}
    \includegraphics[angle=0,width=\columnwidth]{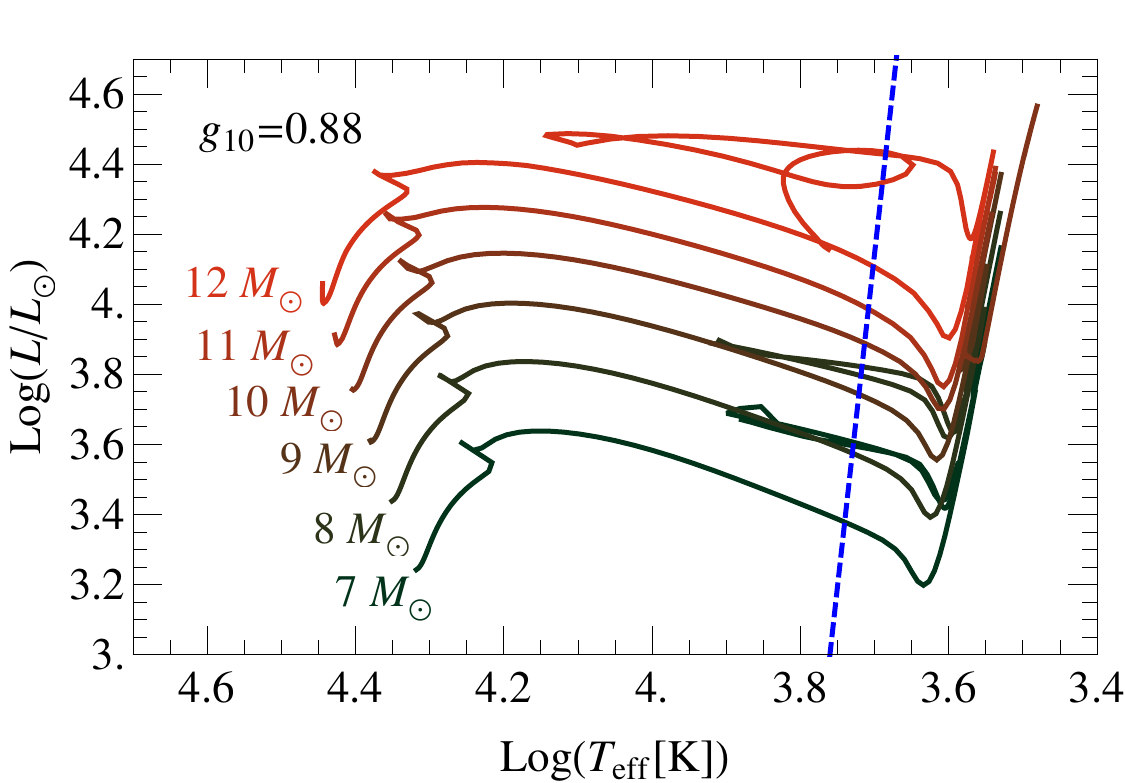}
  \caption{The evolution on the Hertzsprung-Russell (HR) diagram of stars with masses $7-12 M_{\odot}$, with standard cooling \emph{(left)} and with the addition of the $A\gamma\gamma$ coupling at the edge of the sensitivity of CAST \emph{(right)}.  For stars $9 M_{\odot}\lesssim M \lesssim 12 M_{\odot}$ the axion losses completely eliminate the blue loop stage. The dashed lines show the instability strip (conditions for Cepheids).}
  \label{fig:HR}
\end{figure*} 

In this letter, we specialize to the second coupling, $A\gamma\gamma$. In many motivated axion models the dimensionless coefficient $C_{\gamma}$ is ${\cal O}(1)$. For example, for the well-known KSVZ \cite{Kim:1979if,Shifman:1979if} and DFSZ \cite{Dine:1981rt,Zhitnitsky:1980tq} scenarios, we have $|C_{\gamma}|\simeq 1.9$ and $0.7$ respectively. 
It follows that the mass of the axion and the coupling strength to  photons $G_{A\gamma\gamma}=C_{\gamma}\alpha/(2\pi f_{A})$ are proportional to each other. For example, for the KSVZ model one has
\begin{equation}
G_{A\gamma\gamma}^{\rm KSVZ}\simeq 3.7\times 10^{-10}\mbox{ GeV}(m_{A}/1\mbox{ eV}). 
\label{eq:KSVZline}
\end{equation}

For large $f_{A}$ the axion becomes very difficult to detect (``invisible'') and the coupling $G_{A\gamma\gamma}$ becomes one of the key experimental handles~\cite{Sikivie:1983ip}.
Many recent searches have been targeting this coupling~\cite{PDG2010}~\footnote{Or its product with another coupling, as in \cite{Bellini:2012kz}.}, including dark matter detectors, DAMA \cite{Bernabei:2001ny} and CDMS \cite{Ahmed:2009ht}, dedicated axion telescopes, Tokyo \cite{Inoue:2008zp} and CAST\cite{Arik:2008mq}, a reactor experiment, TEXONO \cite{Chang:2006ug}, and even a solar neutrino experiment, Borexino \cite{Bellini:2012kz}. These experiments exclude various segments on the line in Eq.~(\ref{eq:KSVZline}) in the range $10^{0}\lesssim m_{A}\lesssim 10^{5}$ eV, corresponding to the Peccei-Quinn scale $10^{2}$ GeV $\lesssim f_{A}\lesssim 10^{7}$ GeV. 

Remarkably, a stellar cooling bound published over 20 years ago \cite{Raffelt:1987yu} excludes this \emph{entire} range, pushing the bound on $f_{A}$ all the way up to $10^{7}$ GeV.
Given the obvious importance of this result, it is highly desirable to confirm it with more than one type of stellar systems. 
%
This is done below.
The bound of \cite{Raffelt:1987yu} makes use of low mass ($\sim 1.3 M_{\odot}$) stars. 
We show that stars of heavier masses, $8-12M_{\odot}$, can also be used as effective axion probes, an rather unexpected result ({\it cf.} \cite{Raffeltbook}, p.~37). In these stars, axion cooling can \emph{qualitatively} change the evolution, with clear observational consequences. 

We strengthen the astrophysical bound also in another way. For our numerical modeling, we make use of the publicly available and community tested MESA stellar evolution code, to which we release our modifications capturing the axion cooling rates. Our analysis can thus be independently verified and -- we hope -- extended. 

\section{Why Helium Burning?}
The axion-photon coupling leads to energy loss via the Primakoff conversion \cite{Dicus:1978fp,Fukugita:1982gn}: photons convert into axions in the background field of nuclei. The conversion rate is controlled by the finite range of the Coulomb field in plasma, which regulates what would otherwise be a forward scattering logarithmic divergence
\cite{Raffelt:1985nk}. The resulting expression is well established \cite{Raffelt:1985nk,Raffelt:1990yz}; in a nondegenerate medium, per unit mass, the axion loss is
\begin{equation}
\epsilon_{A}= Z(\xi^{2})\frac{G_{A\gamma\gamma}^{2}}{4\pi^{2}} \frac{T^{7}}{ \rho} = 27.2 \frac{\mbox{erg}}{\mbox{g} \cdot \mbox{s} } Z(\xi^{2}) g_{10}^{2} T_{8}^{7} \rho_{3}^{-1}, 
\label{eq:loss}
\end{equation}
where $g_{10}\equiv G_{A\gamma\gamma}/(10^{10}\,\mbox{GeV}^{-1})$, $\rho_3\equiv\rho/(10^3$~g/cm$^{3})$, $T_8 \equiv T/10^8 \rm K$. 
Three powers of temperature come from the photon number density, one from the energy loss per photon, and the remainder from the form of the (plasma-regulated) cross section.

The coefficient $Z(\xi^{2})$ is a function of $\xi^{2}\equiv(\kappa_{S}/2T)^{2}$, with $\kappa_{S}$ being the Debye-Huckel screening wavenumber.
$Z(\xi^{2})$ is given as an integral over the photon distribution (see Eq.~(4.79) in \cite{Raffelt:1990yz}) and is generally ${\cal O}(1)$ for relevant stellar conditions. For example: for the Sun, $\xi^{2}\sim12$ and $Z\sim6$; for the low-mass $He$ burning stars, $\xi^{2}\sim2.5$ and $Z\sim3$ \cite{Raffelt:1990yz}; finally, for a $10 M_{\odot}$  $He$ burning star of interest here, $\xi^{2}\sim0.1$ and $Z\sim0.4$. To include the axion losses in the stellar evolution code, we need a simple, yet accurate, parameterization for $Z(\xi^{2})$. Obviously, this function needs to interpolate between the limits $Z(\xi^{2}\rightarrow0)=(\pi^{3}/30) \xi^{2} \ln(3.99/\xi^{2})$ and $Z(\xi^{2}\rightarrow\infty)=2\pi^{5}/63$, but the interpolation needs to also accurately reproduce the intermediate regime, since the physically interesting values of $\xi^{2}$ lie there. We propose using
\begin{equation}
\label{eq:alternativefit}
{\textstyle Z(\xi^{2})\simeq 
\left(\frac{1.037\xi^{2}}{1.01+\xi^{2}/5.4}+\frac{1.037\xi^{2}}{44+0.628\xi^{2}}\right)
\ln\left(3.85+\frac{3.99}{\xi^{2}}\right)}.
\end{equation}
The accuracy of this parameterization is better than 2\% over the entire range of $\xi$.

Using the cooling rate in Eqs.~(\ref{eq:loss},\ref{eq:alternativefit}), we plot in Fig.~\ref{fig:RhoThistories} the region where the axion cooling with $g_{10}=1$ comprises at least 90\% of the overall non-photon energy loss. The effect of the axion is pronounced at moderate temperatures and densities, ordinarily the domain of photoproduction ($\gamma e^{-}\rightarrow e^{-}\nu\bar\nu$); for higher temperatures, it is overtaken by the SM pair production ($e^{+}e^{-}\rightarrow\nu\bar\nu$), while for higher densities, the SM plasmon decay dominates ({\it cf.} \cite{Raffelt:1985nk,Heger:2008er}). Since the rate of axion emission increases with temperature (as illustrated by the two dashed isocontours), the optimal temperatures for probing Primakoff losses are generally in the upper part of the shaded region, $1\times10^{8}\mbox{ K}\lesssim T_{C}\lesssim4\times 10^{8}$ K. These are precisely the conditions at which Helium burns.
For illustration, we show two curves depicting the evolution of the central temperature and density in 1 $M_{\odot}$ and 15 $M_{\odot}$ stars. The calculations were carried out with the MESA code, without the axion cooling. Next, we show what happens as this cooling is added.

\begin{figure*}[ht]
  \includegraphics[angle=0,width=0.57\textwidth]{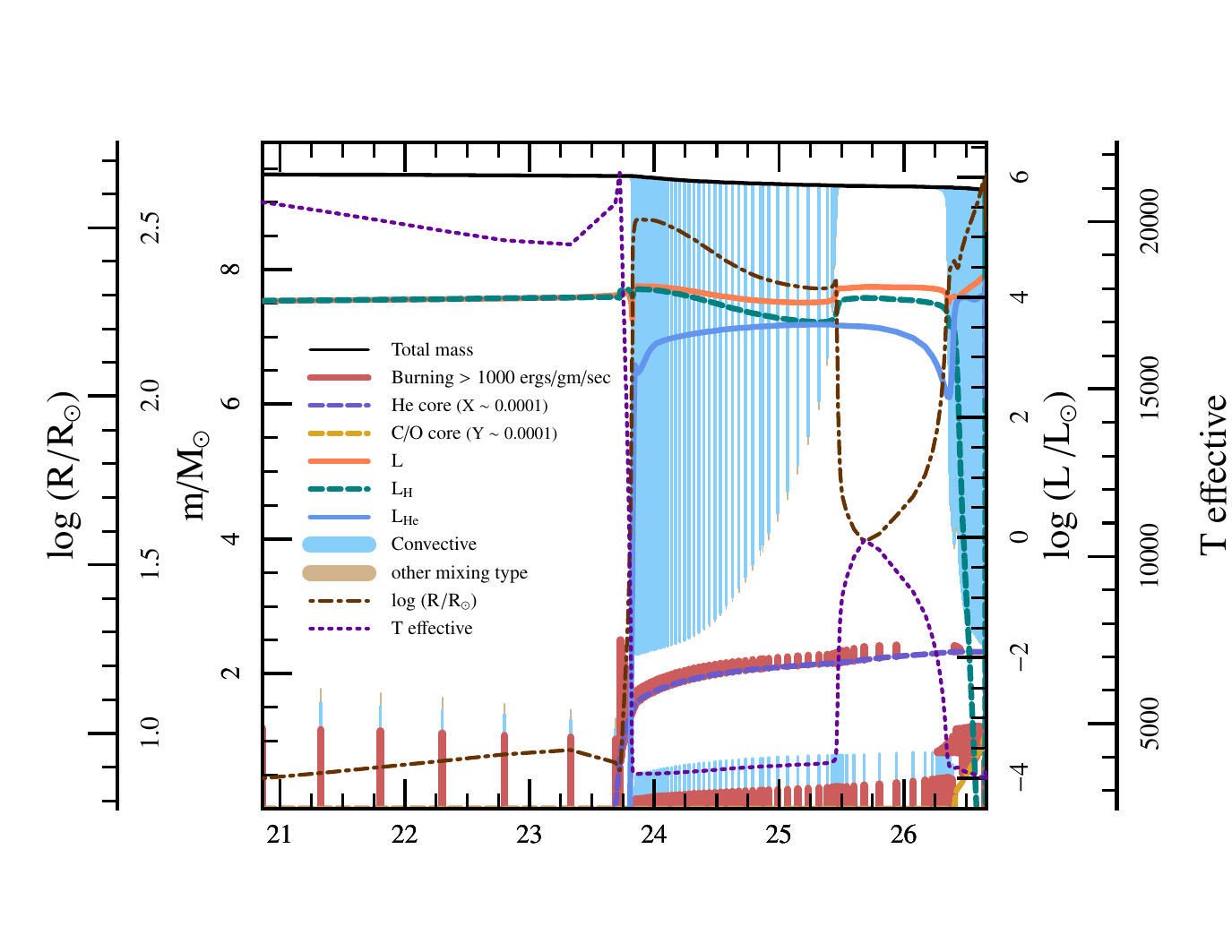}
    \includegraphics[angle=0,width=0.57\textwidth]{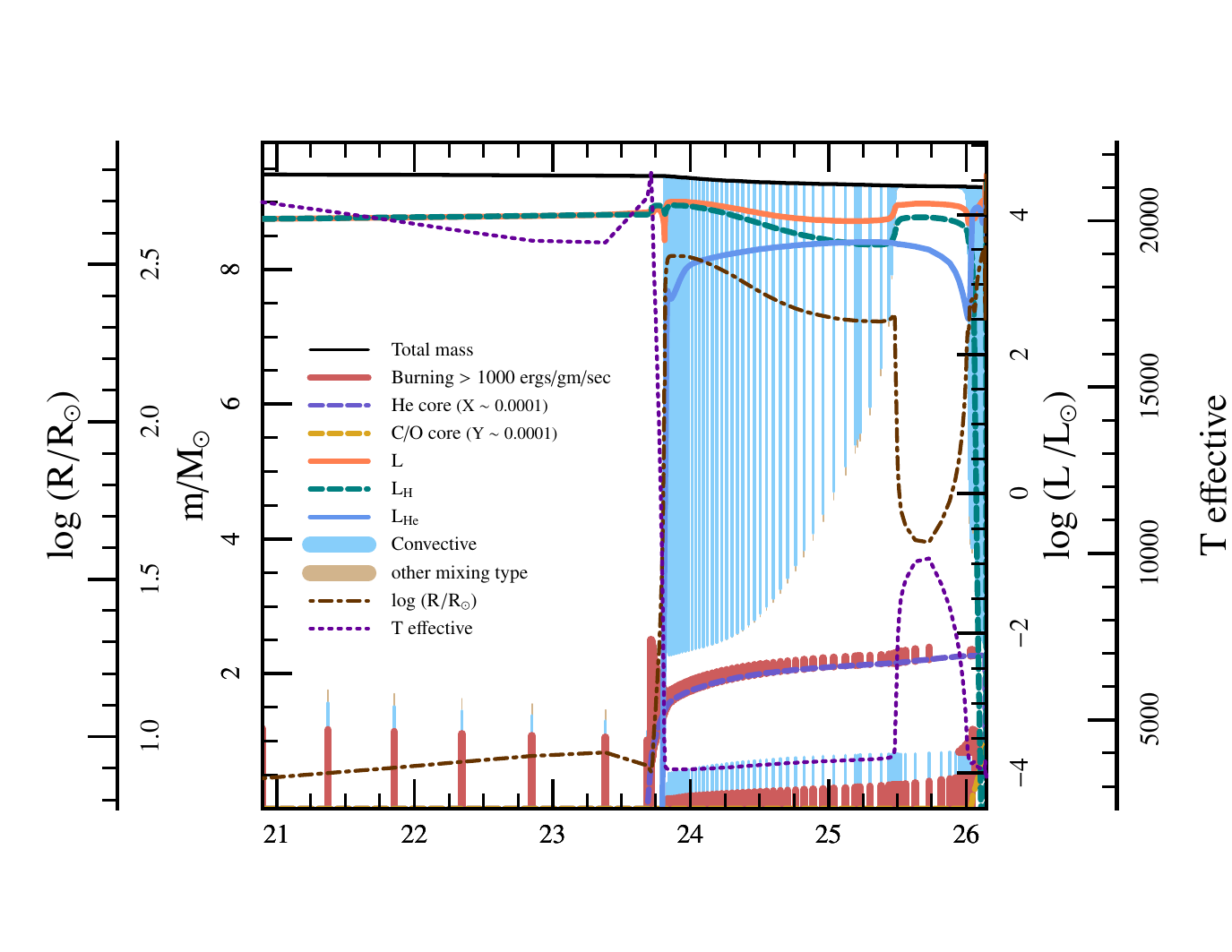}
   \includegraphics[angle=0,width=0.57\textwidth]{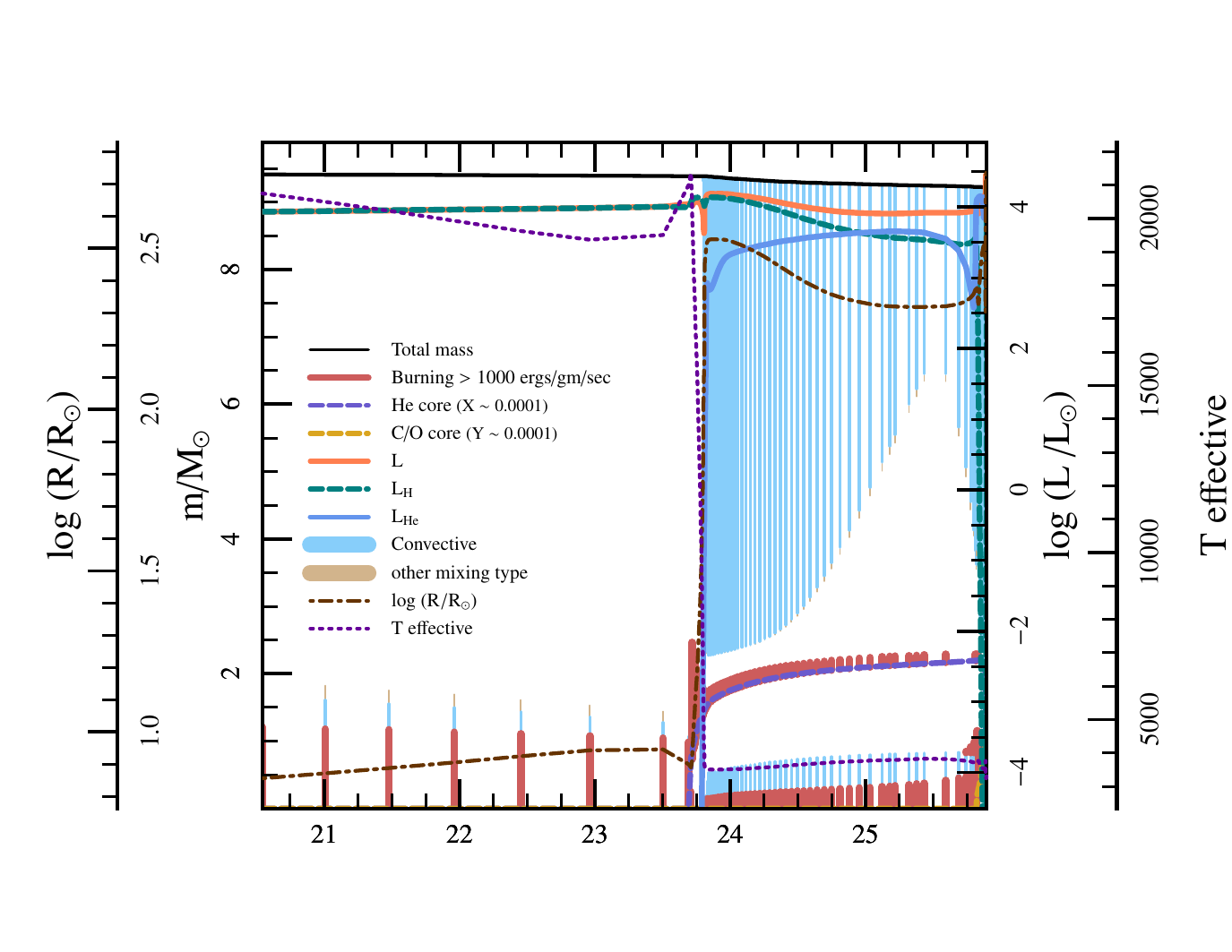}
  \caption{Impact of the $A\gamma\gamma$ coupling on the evolution of a 9.5 $\Msun$ star through the $He$ burning stage. Shown are the cases $g_{10}=0$ (\textit{top}), $g_{10}=0.6$ (\textit{middle}), and $g_{10}=0.8$ (\textit{bottom}). The horizontal axis shows the age of the star, in $10^{6}$ years. Axion losses speed up $He$ burning, resulting in the shortening and eventual elimination of the blue loop stage.}
  \label{fig:Internal_structure}
\end{figure*}

\section{Axion Cooling in MESA evolution code}
MESA (``Modules for Experiments in Stellar Astrophysics'') is a robust, \emph{open source}, modular 1D stellar evolution package~\cite{mesapage}. Its release several years ago represents a very  significant development in the field of stellar astrophysics. The ``instrument paper'' \cite{Paxton:2010ji} has shown MESA to be remarkably versatile, capable of handling not only solar-mass stars, but also objects as diverse as $10^{-2}M_{\odot}$ and $10^{3}M_{\odot}$. 

MESA has been widely accepted by the astrophysics community. Yet, its applications to studying new physics have so far been confined to models of modified gravity~\cite{Davis:2011qf,Jain:2012tn,Chang:2010xh}. To the best of our knowledge, the present letter represents the first use of MESA to constraining new particles. Given its openness and robustness, we are convinced that over time MESA will become a standard tool for probing fundamental physical processes in stars.

We added the axion cooling effect, Eqs.~(\ref{eq:loss},\ref{eq:alternativefit}) to the MESA code (specifically, to the neutrino energy loss routine \verb|neu.f|). Our modified code is being made publicly available~\cite{ourpage}, so that our analysis can be reproduced and further extended.
As a simple verification test, we ran a 1.3 $M_{\odot}$ model with and without the axion cooling, paying particular attention to the duration of the $He$ burning (the Horizontal Branch, HB, stage). This is the model studied in \cite{Raffelt:1987yu} and which has been the basis for the published bounds over the last two decades. The model ran straightforwardly through Hydrogen burning, Helium flash, and the HB stage. The resulting durations of the HB stage were in excellent agreement with \cite{Raffelt:1987yu}: $1.2\times10^8$ yrs without the axion $0.7\times 10^8$ yrs upon adding axion-photon coupling with $g_{10}=1$ (see also  \cite{Raffeltbook}, page 81). Notice that in this case there is no \emph{qualitative} change of the evolution. The argument constraining the axion is based on counts of low-mass HB stars in stellar clusters and in the galactic disk. Faster burning of $He$ due to axion losses would reduce the counts.

We next turn to our main calculation, the impact of the axion on stars of masses $\sim7-12M_{\odot}$. The results are shown in Fig.~\ref{fig:HR}, as the evolutionary tracks for these stars in the Hertzsprung-Russell (HR) diagram for $g_{10}=0$ and $g_{10}=0.88$. The second value represents the limit of sensitivity of the CAST experiment for very small axion mass (off the KSVZ line). We see that even such small axion coupling \emph{qualitatively} changes the evolution. Normally, these stars, after reaching the red giant tip, travel back to the left (blue) side of the HR diagram. This is the well-known \emph{blue loop} phenomenon \cite{Hayashi1962,Lauterborn1971a,Lauterborn1971b,kippenhahn,XuLi1}. With the axion cooling, however, for stars with $9M_{\odot}\lesssim  M\lesssim 12 M_{\odot}$, this evolutionary stage disappears altogether.

\section{Discussion}
Let us examine the physics behind the disappearance of the blue loop. It is helpful to look at the evolution of the internal structure through the $He$ burning stage. Fig.~\ref{fig:Internal_structure} shows this evolution for a representative $9.5M_{\odot}$ star, with the horizontal axis showing the stellar age in millions of years (Myr). The top panel corresponds to the standard case ($g_{10}=0$), while the middle and the bottom one have $g_{10}=0.6$ and $0.8$ respectively. 

In all three models, Helium is ignited at $23.8$ Myr and the preceding evolution is not noticeably changed by axion losses. The axion losses, however, do speed up the $He$ burning stage, as expected (more losses require faster burning). The duration of this stage decreases from $\sim2.6$ Myr for $g_{10}=0$ to $\sim2$ Myr for $g_{10}=0.8$. Notice that in the top panel, at 25.5 Myr the star undergoes a transformation: its radius contracts, while its surface temperature rises. This is the blue loop phenomenon: the star transitions from a red giant with a large convective envelop to a more compact blue giant with a radiative envelop. The same transition also occurs in the middle panel, but notice that the modest overall shortening of the $He$ burning stage significantly shrinks the blue loop stage. In the bottom panel, the $He$ burning stage is shortened enough that the blue loop simply does not have time to start. The core exhausts its Helium, then contracts to Carbon ignition ({\it cf.} Fig.~\ref{fig:RhoThistories}), at which point SM neutrino losses increase so much that the rest of the evolution proceeds in a very short time (see, {\it e.g.}, \cite{Woosley:2006ie}). 

This is the basis of our argument for constraining the axion-photon coupling: a \emph{quantitative} change -- speed-up of $He$ burning -- for these stars leads to a \emph{qualitative} change in the evolution -- elimination of the blue loop. The elimination of the loop would have at least two obvious signatures. (i) An entire observed population of stars, blue $He$ burning giants, would not be accounted for. Detailed observations of blue loop populations exist (see, {\it e.g.}, \cite{Skillman:2002aa} and \cite{McQuinn:2011bb}, particularly Fig.~3 therein). (ii) As stars go through a blue loop, they cross the Instability Strip and become Cepheid variables. Without the blue loop \cite{kippenhahn}, one cannot account for the existence of Cepheid stars with the broad range of pulsation periods (corrseponding to $\sim8-11\Msun$). The initial crossing of this strip, as the star adjusts from its main sequence configuration to a $He$ burning red giant state, proceeds too fast to give large enough numbers of these variable stars.  

Our investigations so far show that the resulting bound is somewhere between a rather conservative $g_{10}\lesssim0.8$ and most aggressive $g_{10}\lesssim0.5$. The exact value depends on the detailed analysis of the observations and the physics of the simulation. While such a detailed study is well beyond the scope of the present letter, below we summarize several relevant considerations.

First, for our bound we require a complete disappearance of the blue loop, eliminating the entire observed population of stars. This is a conservative requirement. Given accurate counts, it may be possible to check whether the number of stars in the blue loop phase is \emph{reduced}. For example, in Fig.~\ref{fig:Internal_structure} the middle panel shows that $g_{10}=0.6$ would reduce the time a $9.5M_{\odot}$ star spends on the blue loop by a factor of two. (Notice, for comparison, that to get the same sensitivity for $g_{10}$ from solar-mass stars requires knowing the numbers of HB stars to a $\sim 10$\% precision~\cite{Raffeltbook}.)

Second, one can consider the effect of the axion on stars of different masses and find which stars have the most sensitivity. In our investigations, for example, we found that for $10.5 M_{\odot}$ stars the blue loop disappeared already for $g_{10}\sim 0.5$. The observational signature in this case could be a gap in the observed periods of Cepheid stars, which vary as a function of stellar mass \footnote{We thank several members of the OSU Astronomy Department for this important suggestion.}. Again, our bound is conservative with respect to this point.

Third, the details of the blue loop depend on the treatment of the convection physics in the code \cite{eid} \footnote{For more massive stars, the mass loss phenomenon becomes another important astrophysical uncertainty. This is the reason why we select stars $\le12\Msun$ for the present study.}. In our investigations with MESA we confirm that varying, e.g., the mixing length parameter shifts the exact value of $g_{10}$ at which the blue loop disappears. Understanding stellar convection is presently a focus of an active effort in the stellar astrophysics community. Since our code is being made public, we invite the members of this community to test the impact of various convection prescriptions -- and other physical assumptions and numerical methods -- on the axion bound. 
We hope, in time, this will result in a stronger bound on the axion. Tentatively, here we choose to state the conservative bound, $g_{10}\lesssim 0.8$.

\section{Conclusions}
We have obtained a new astrophysical bound on the axion-photon interactions, by considering the evolution of stars $\sim7-12$ times more massive than the Sun. The sensitivity of these stars to the axion-photon coupling compares favorably to the published bound $g_{10}<1$ from the solar mass stars  \cite{PDG2010} \footnote{Our bound also extends to somewhat heavier axion masses, since the $He$ burning cores of $10M_{\odot}$ stars are $\sim 30$\% hotter than those of the solar mass stars.}. Sufficiently large axion-photon coupling is shown to eliminate the blue loop stage of the evolution, leaving one without an explanation for the existence of Cepheid stars in a broad range of pulsation periods. This is the second time massive stars are used to constrain particle physics beyond the Standard Model and, as in the case of neutrino magnetic moment \cite{Heger:2008er}, axion is also capable of qualitatively changing the stellar evolution.


\begin{acknowledgments}
We would like to thank Bill Paxton for leading the development of the MESA code and for his quick responses to our queries. We also gladly acknowledge helpful discussions with Casey Meakin at LANL and with several members of the OSU Astronomy Department. This research was supported at LANL by the DOE Office of Science and the LDRD Program.
\end{acknowledgments}


\bibliography{axionbib}

\end{document}